\documentclass[prb,twocolumn,superscriptaddress,showpacs,showkeys]{revtex4}

\usepackage{graphics}
\usepackage{color}
\usepackage{dcolumn}
\usepackage{amsmath}
\usepackage{amssymb}
\usepackage[colorlinks=false]{hyperref}

\newcommand{\ie}{{\it i.e.}}
\newcommand{\eg}{{\it e.g.}}

\newcommand{\etc}{{\it etc.}}

\newcommand{\YBCO}{YBa$_2$Cu$_3$O$_7$}

\newcommand{\state}[1]{$\stat{#1}$}

\newcommand{\stat}[1]{%
  \begingroup
  \def\delimiter##1##2##3##4##5##6##7##8{##10##3##4##5}%
  \mathcode`\0=\downarrow \mathcode`\1=\uparrow
  \mathcode`\d=\downarrow \mathcode`\u=\uparrow
  \mathcode`\D=\Downarrow \mathcode`\U=\Uparrow
  \mathcode`\-=\circ
  #1\relax
  \endgroup
}
\let\stt\stat
\renewcommand{\stat}[1]{\protect\stt{#1}}

\begin{document}

\title{
  Fluxon-semifluxon interaction in an annular long Josephson 0-$\pi$-junction
}

\author{E.~Goldobin}
\email{gold@uni-tuebingen.de}
\homepage{http://www.geocities.com/e_goldobin}
\affiliation{
  Physikalisches Institut -- Experimentalphysik II,
  Universit\"at T\"ubingen,
  Auf der Morgenstelle 14,
  D-72076 T\"ubingen, Germany
}

\author{N.~Stefanakis}
\affiliation{
  Institut f\"ur Theoretische Physik,
  Universit\"at T\"ubingen,
  Auf der Morgenstelle 14,
  D-72076 T\"ubingen, Germany
}

\author{D.~Koelle}
\author{R.~Kleiner}
\affiliation{
  Physikalisches Institut -- Experimentalphysik II,
  Universit\"at T\"ubingen,
  Auf der Morgenstelle 14,
  D-72076 T\"ubingen, Germany
}

\pacs{
  74.50.+r,   
  85.25.Cp    
  74.20.Rp    
}

\keywords{
  Long annular Josephson junction, sine-Gordon,
  half-integer flux quantum, semifluxon,
  0-pi-junction
}



\date{\today}

\begin{abstract}
  We investigate theoretically the interaction between integer and half-integer Josephson vortices (fluxons and semifluxons) in an annular Josephson junction. Semifluxons usually appear at the 0-$\pi$-boundary where there is a $\pi$-discontinuity of the Josephson phase. We study the simplest, but the most interesting case of one $\pi$-discontinuity in a loop, which can be created only artificially. We show that measuring the current-voltage characteristic after injection of an integer  fluxon, one can determine the polarity of a semifluxon. Depending on the relative polarity of fluxon and semifluxon the static configuration may be stable or unstable, but in the dynamic state both configurations are stable. We also calculate the depinning current of $N$ fluxons pinned by an arbitrary fractional vortex.
\end{abstract}


\maketitle

\section{Introduction}
\label{Sec:Intro}

For conventional Josephson junctions the first Josephson relation reads $I_s=I_c\sin(\phi)$, where $I_s$ is the supercurrent through the junction, $I_c$ is the critical current and $\phi$ is the so-called Josephson phase which is equal to the difference of the phases of the quantum-mechanical macroscopic wave functions in the electrodes. The Josephson relation for a Josephson $\pi$-junction is $I_s=-I_c\sin(\phi)=I_c\sin(\phi+\pi)$, \ie, a $\pi$-junction can be considered as a junction with negative critical current or having an additional phase shift of $\pi$ between the phases of the wave functions (therefore the name). Accordingly, conventional Josephson junctions are sometimes called 0-junctions.

If one considers a 1D long Josephson 0-$\pi$-junction (LJJ) made of alternating parts with positive and negative critical currents (0 and $\pi$-parts), half-integer flux quanta (semifluxons\cite{Goldobin:SF-Shape,Xu:SF-shape}) may spontaneously form at the boundaries between 0 and $\pi$ regions. 

Semifluxons are very interesting objects which are not yet studied in detail, first of all because up to now it was rather difficult to fabricate 0-$\pi$-junctions. Recently several groups succeeded to demonstrate 0-$\pi$-junctions based on various technologies: \YBCO-Nb ramp zigzags\cite{Smilde:ZigzagPRL,Hilgenkamp:zigzag:SF}, grain boundary junctions based on tri- and tetra-crystals\cite{Chesca:2002:YBCO4Crystal-pi-SQUID,Chesca:2003:LCCO4Crystal} and Nb junctions based on an artificially created discontinuity\cite{Goldobin:Art-0-pi}. Both 0 and $\pi$ junctions from the Superconductor-Ferromagnet-Superconductor family were demonstrated by several groups\cite{Kontos:2002:SIFS-PiJJ,Ryazanov:2001:SFS-PiJJ,Blum:2002:IcOscillations} but a 0-$\pi$-LJJ was not reported yet. Semifluxons were observed using SQUID microscopy in different types of 0-$\pi$-LJJs \cite{Hilgenkamp:zigzag:SF,Kirtley:SF:T-dep,Kirtley:SF:HTSGB,Mints:2002:SplinteredVortices@GB}.

A single semifluxon, formed in a LJJ of length $L\gg\lambda_J$ with one 0-$\pi$-boundary, is pinned at this 0-$\pi$-boundary\cite{Goldobin:SF-ReArrange,Susanto:SF-gamma_c} and can have positive or negative polarity carrying the flux $+\Phi_0/2$ or $-\Phi_0/2$, respectively. The bias current from 0 up to $\frac{2}{\pi}I_c$ ($I_c=j_c w L$ is the ``intrinsic'' critical current, $w$ is the junction width) cannot move the semifluxon but just changes its shape\cite{Goldobin:SF-ReArrange}. This property suggests to use semifluxons in information storage devices, classical or quantum. 

In the classical regime the polarity of a semifluxon (positive or negative) will encode a logical 0 or 1. The information encoding using semifluxons is somewhat more robust than using fluxons because semifluxons cannot ``escape'' as they are pinned at the discontinuity point. The switching between these states can be done by injecting a single flux quantum of proper polarity into the junction. 

In the quantum limit a semifluxon having two possible polarities is similar to a spin with two possible orientations (up or down). Therefore we often use spin notation to denote the polarity of the semifluxon, \ie, \state{u} or \state{d}. Analogously, we denote the polarity of a fluxon as \state{U} or \state{D}. It seems that a semifluxon is an interesting candidate to realize a qubit. It still remains a challenging task to find out whether the semifluxon may stay in the superposition of both states and perform quantum tunneling between them or not. In comparison with a fluxon based qubit the one based on a semifluxon should be more robust as both \state{u} and \state{d} states represent the ground state of the system, while the fluxon of any polarity is an excited state, the ground state being the one with constant phase.

For both quantum and classical bits one needs a way to determine the final state  of the semifluxon, \ie, read out its polarity. Imagine the simplest situation: a single semifluxon of unknown polarity in an annular LJJ
\footnote{\label{Ft:OnePt}We note that such annular junctions cannot be constructed using natural 0 or $\pi$ junctions because one part cannot be simultaneously 0 and $\pi$. There should be always an even number of 0-$\pi$-joints. In an experiment the proposed situation can be realized using artificial 0-$\pi$-LJJ based on conventional superconductors with a pair of $\delta$-injectors\cite{Goldobin:Art-0-pi}. This case of only one discontinuity in a ring is very interesting as it allows to go beyond the possibilities offered by nature.}. Let us inject a fluxon of a certain polarity into this Josephson ring somewhere far from the semifluxon position, \eg, using a pair of current injectors\cite{Ustinov:2002:ALJJ:InsFluxon}. If the polarities of the fluxon and of the semifluxon are different, the ``annihilation'' between fluxon and semifluxon will result in a pinned semifluxon of the opposite polarity (we assume that the bias current is zero during fluxon injection). The $I$--$V$ characteristic (IVC) of the resulting state has a rather large maximum supercurrent (depinning current of a semifluxon) $\gamma_c^\uparrow=2/\pi$ in normalized units ($\gamma=I/I_c$). On the other hand, if the fluxon and the semifluxon are of the same polarity, no ``annihilation'' takes place. If the bias current is applied, the fluxon starts moving passing through the semifluxon resulting in a finite voltage across the LJJ and rather low $\gamma_c$. In this case the depinning current $\gamma_c^{\stat{Uu}}$ is determined by the repulsion force between fluxon and semifluxon. The fluxon should overcome this pinning by the semifluxon to start moving around the ring.

In this paper we propose a technique to test the polarity of semifluxons by introducing test fluxons into the Josephson ring. Measuring the current voltage characteristic we can determine the polarity of semifluxon(s) before the fluxon was introduced. We also study resulting states which have different critical (depinning) currents and different dynamics.

In section \ref{Sec:Model} we introduce the model which is used for numerical simulations. The numerical results are presented and discussed in section \ref{Sec:SimRes}. Section \ref{Sec:Conclusion} concludes this work.

\section{The Model}
\label{Sec:Model}

The dynamics of the Josephson phase in a LJJ consisting of alternating 0 and $\pi$ parts can be described by the 1D perturbed sine-Gordon equation\cite{Goldobin:SF-Shape}
\begin{equation}
  \phi_{xx}-\phi_{tt}-\sin\phi = \alpha\phi_t-\gamma-\theta_{xx}(x)
  , \label{Eq:sG-phi}
\end{equation}
where $\phi(x,t)$ is the Josephson phase and subscripts $x$ and $t$ denote the derivatives with respect to coordinate $x$ and time $t$. In Eq.~(\ref{Eq:sG-phi}) the spatial coordinate is normalized to the Josephson penetration depth $\lambda_J$ and the time is normalized to the inverse plasma frequency $\omega_p^{-1}$; $\alpha=1/\sqrt{\beta_c}$ is the dimensionless damping ($\beta_c$ is the McCumber-Stewart parameter); $\gamma=j/j_c$ is the external bias current density normalized to the critical current density of the junction. The function $\theta(x)$ is a step function which is $\pi$-discontinuous at all points where 0 and $\pi$ parts join and is a constant equal to $\pi n$ within each part ($n$ is an integer). For example, $\theta(x)$ can be equal to zero along all 0-parts and $\pi$ along all $\pi$-parts.

It is clear from Eq.~(\ref{Eq:sG-phi}) that $\phi(x)$ is also $\pi$-discontinuous at the same points as $\theta(x)$. Therefore, we often call the points where 0 and $\pi$ parts join \emph{phase discontinuity points}.

Note that to describe 0-$\pi$-LJJs other authors\cite{Xu:SF-shape,Kirtley:IcH-PiLJJ,Buzdin:2003:phi-LJJ} often use directly the equation with alternating critical current density written for the continuous phase $\mu(x,t)$
\begin{equation}
  \mu_{xx}-\mu_{tt}\pm\sin\mu = \alpha\mu_t-\gamma(x)
  . \label{Eq:sG-mu}
\end{equation}
The Eqs.~(\ref{Eq:sG-phi}) and (\ref{Eq:sG-mu}) are, actually, equivalent and one can be obtained from the other by substitution $\phi(x,t)=\mu(x,t)+\theta(x)$\cite{Goldobin:SF-Shape}.

In case of an annular LJJ one should use periodic boundary conditions (b.c.) to solve Eq.~(\ref{Eq:sG-phi}) or (\ref{Eq:sG-mu}). In the case of a conventional annular LJJ without phase discontinuities the boundary conditions are expressed as follows 
\begin{equation}
  \phi(L,t)=\phi(0,t)\pm 2\pi n_F,
  \label{Eq:BC:Conv}
\end{equation}
where $n_F$ is the number of flux quanta (Josephson vortices) trapped in the ring. Note that when there are no discontinuities $\mu\equiv\phi$, so that b.c.~(\ref{Eq:BC:Conv}) holds for $\mu$ also.

For the case of an annular LJJ with discontinuities the boundary conditions for $\phi$ are still given by Eq.~(\ref{Eq:BC:Conv}). This can be understood using the following \textit{gedanken} experiment. Imagine that we start from the state without discontinuities $\phi(x)=0$. Then we slowly increase a discontinuity, \eg, by using a pair of closely located $\delta$-injectors\cite{Goldobin:Art-0-pi}, at some point $x=x_0$ from the value 0 to some value $\kappa$. It is clear that the Josephson phase $\phi(x)$ changes somehow on the length scale $\lambda_J$ in the vicinity of $x_0$ to compensate (to react on) this discontinuity, \eg, by forming a fractional vortex with the center at $x=x_0$. In any case the phase $\phi(x)$ and its derivative $\phi_x(x)$ are smooth and continuous functions all along the junction except for the point $x=x_0$. Assuming that the discontinuity point does not coincide with $x=0$ or with $x=L$, we can write $\phi(0)=\phi(L)$ and $\phi_x(0)=\phi_x(L)$. In the presence of additional fluxons trapped in the junction we get b.c. (\ref{Eq:BC:Conv}) even in the presence of the discontinuity points.

The b.c. for $\mu$ can be written, recalling that $\mu=\phi-\theta(x)$
\begin{equation}
  \mu(L,t)=\mu(0,t)+ \pi n_{\rm SF} + 2\pi n_F,
  \label{Eq:BC:Unconv}
\end{equation}
where $n_{\rm SF}$ is the sum of all discontinuities (semifluxons) in the ring. Note that if
\begin{equation}
  \theta(x) = \sum_{i=1}^N \kappa_i {\cal H}(x-x_i)
  , \label{Eq:theta_ex}
\end{equation}
where $\kappa_i$ is the $i$-th discontinuity and ${\cal H}(x)$ is a Heavyside step function, the expression for $n_{\rm SF}$ is
\begin{equation}
  n_{\rm SF} = \sum_{i=1}^N \kappa_i
  , \label{Eq:n_SF}
\end{equation}

In this paper we will investigate the following two cases 
\begin{enumerate}
  \item A fluxon is injected into the annular LJJ containing a negative semifluxon, \ie,  $n_F=+1$, $n_{\rm SF}=-1$.
  \item A fluxon is injected into the annular LJJ containing a positive semifluxon, \ie, $n_F=+1$, $n_{\rm SF}=+1$.
\end{enumerate}

\section{Numerical results}
\label{Sec:SimRes}

The simulations were performed using \textsc{StkJJ} software\cite{StkJJ} and were confirmed by an independently written program\footnote{written by N. Stefanakis}.

We use $\alpha=0.1$ for all results reported here. This value is not very high, so that it allows to observe some dynamics. On the other hand, it is not very low as the majority of 0-$\pi$-junctions has rather high damping. In the case $\alpha\agt1$ the static results related to the reading out the state of a semifluxon hold, but no dynamical effects such as fluxon steps can be observed.

In order to visualize and to understand the fluxon and semifluxon dynamics we plot their trajectories on the $(x,t)$ plane. Usually, to track the trajectory of a fluxon one tracks the trajectory of its center. Since at a given instant of time the phase $\mu(x)$, corresponding to the fluxon solution, changes from $0$ at $x\to-\infty$ to $2\pi$ at $x\to+\infty$, it is assumed that the center of a fluxon coincides with the point where the phase $\mu=\pi$. Since the phase is defined modulo $2\pi$, in the general case the center of fluxons is situated at points where $\mu=\pi+2\pi k$ with integer $k$. To distinguish between fluxons and antifluxons, we also check the sign of the phase derivative $\mu_x$ (magnetic field) at the point where $\mu=\pi+2\pi k$. If the sign is positive, then it is a fluxon and we plot its position as a black point on $x$--$t$ plane. If $\mu_x<0$, then we plot it as a gray points on the $x$--$t$ plane.

When we deal with semifluxons, the idea is the same, but since the phase of a semifluxon changes from 0 to $\pi$ ($\mod \pi$) we have to define the centers of semifluxons as the points where $\mu=\frac{\pi}{2}+\pi k$.
The sign of $\mu_x$ at the points where $\mu=\frac{\pi}{2}+\pi k$ is used to distinguish between semifluxons of positive and negative polarity.
With this definition, every $2\pi$-fluxon carrying the integer flux $\Phi_0$ results in two points on the $x$--$t$ plane, \ie, its trajectory will be represented by a double line.

Below we present numerical results obtained for an annular junction of length $L=8\lambda_J$.

\subsection{One semifluxon in a ring}

First we investigate numerically an annular LJJ with only one phase discontinuity point.

Here and below, without loosing generality, we assume that the injected fluxons have positive polarity \state{U} [$n_F=+1$ in (\ref{Eq:BC:Unconv})]. 

\begin{figure*}[tb]
  \centering\includegraphics{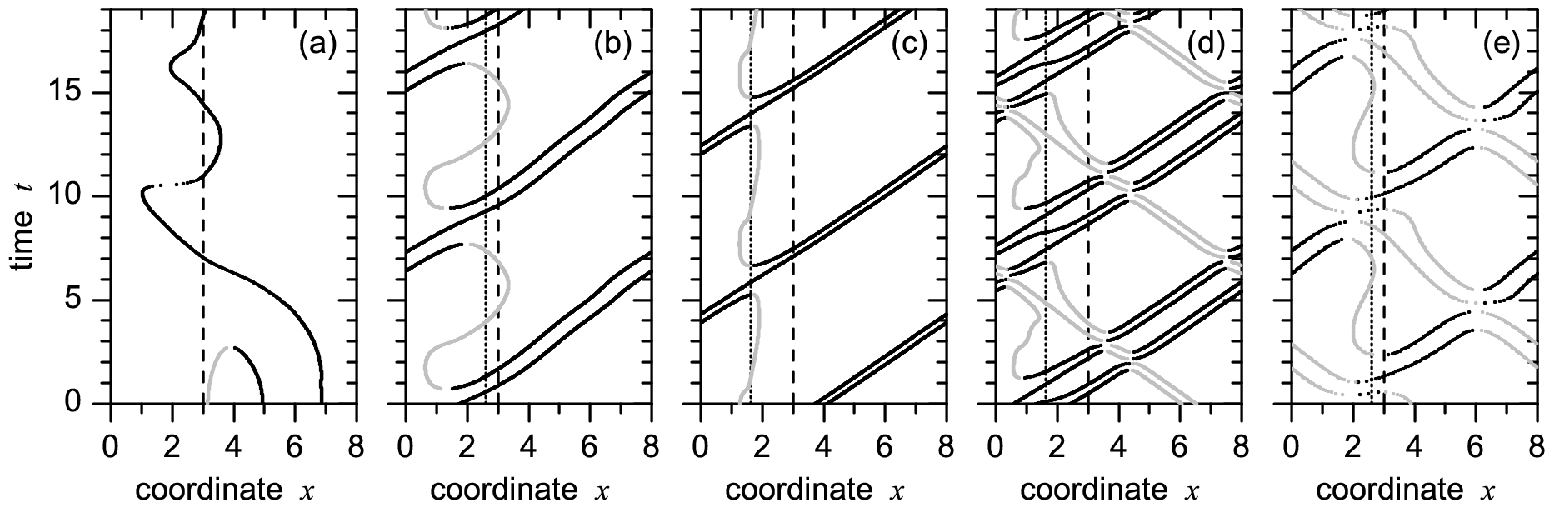}
  \caption{
    Trajectories of fluxons (double line) and semifluxons (single line)  corresponding to annihilation $\stat{U}+\stat{d}=\stat{u}$ at $\gamma=0$ (a); dynamics in the state \state{Ud} at the first fluxon step at $\gamma=0.3$ (b); at $\gamma=0.6$ (c); dynamics at the third fluxon step (state \state{UUDu}) at $\gamma=0.6$ (d); dynamics at the second fluxon step (state \state{UDu}) at $\gamma=0.3$. The vertical dashed line shows the position of the discontinuity. The vertical dotted line shows the position of the center of the static semifluxon at the corresponding bias. In (a) for $\gamma=0$ both lines coincide.
  }
  \label{Fig:Traj:Ud}
\end{figure*}
\newcommand{\FigIVCUdltZeroBias}{solid black symbols}
\newcommand{\FigIVCUdltFiniteBias}{open symbols}
\newcommand{\FigIVCuUD}{solid gray symbols}
\begin{figure}[tb]
  \centering\includegraphics{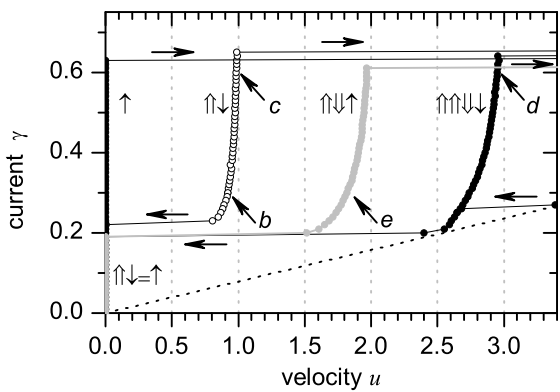}
  \caption{
    IVC after injection of a fluxon into a 0-$\pi$-LJJ containing a semifluxon of negative polarity. The \FigIVCUdltZeroBias{} show the IVC after annihilation, \FigIVCUdltFiniteBias{} show the IVC which can be traced if injection takes place at finite $2/\pi>\gamma>\gamma_r\approx 0.21$. Dotted line shows the position of the McCumber branch (uniform phase-whirling state).
  }
  \label{Fig:IVC:Ud}
\end{figure}

\subsubsection{Negative semifluxon}

If initially the semifluxon has negative polarity \state{d} [$n_{\rm SF}=-1$ in (\ref{Eq:BC:Unconv})] the fluxon is attracted by the semifluxon and, in the absence of a bias current, they ``annihilate'', resulting in a positive semifluxon \state{u}. 

In Fig.~\ref{Fig:Traj:Ud}(a) one can see the ``annihilation'' process of the semifluxon situated at $x=3$ and the fluxon injected at $x=6$ (it corresponds to two lines at $x\approx5$ and $x\approx7$ at $t=0$) for zero bias current $\gamma=0$. If we trace the IVC of the state after annihilation, we get the curve shown in Fig.~\ref{Fig:IVC:Ud} by \FigIVCUdltZeroBias. As a horizontal axis we use the fluxon velocity $u$ normalized to the Swihart velocity and proportional to the voltage across the junction. In this notation a single fluxon has an asymptotic velocity 1, two fluxons have $u\to2$ as $\gamma$ grows, \etc. At $u=0$ the state of the system is the semifluxon \state{u} with maximum supercurrent (semifluxon depinning current) $\gamma_c^\uparrow=2/\pi\approx0.63$\cite{Kato:1997:QuTunnel0pi0JJ,Goldobin:SF-ReArrange,Zenchuk:2003:AnalXover,Ustinov:2002:ALJJ:Ic(Iinj)}. When the bias current exceeds this value the system switches to the McCumber branch. By sweeping the bias current back one may trace the step with asymptotic velocity $u=3$, corresponding to the state \state{UUDd}, which is formed probably due to the topological instability of the solution similar to the formation of zero-field steps in conventional LJJ. The trajectories corresponding to the bias point $d$ can be seen in Fig.~\ref{Fig:Traj:Ud}(d).

What happens if we inject a fluxon while having a non-zero bias current? 
If the bias current as well as the distance between fluxon injection point and semifluxon are large enough, the fluxon approaches the semifluxon with rather high velocity and annihilation will not take place --- the fluxon will simply pass through the semifluxon. Thus, injecting a fluxon at a finite bias current we can trace an IVC which is shown in Fig.~\ref{Fig:IVC:Ud} by \FigIVCUdltFiniteBias. The semifluxon trajectories at the bias points $b$ and $c$ are shown in Fig.~\ref{Fig:Traj:Ud}(b)--(c), accordingly. One can see that in average a semifluxon is shifted by the bias current away from the discontinuity point and oscillates around this new equilibrium position as the fluxon bumps it. Comparing Fig.~\ref{Fig:Traj:Ud}(b) and (c), one can also notice that the fluxon's double line is more tight in (c) which is a result of relativistic contraction. Note that after the bias current is reduced to zero the annihilation takes place anyway and the system returns to the IVC corresponding to the state \state{u}, without a moving fluxon, so that further sweeping of $\gamma$ shows only the curve drawn by \FigIVCUdltZeroBias{} in Fig.~\ref{Fig:IVC:Ud}.

Actually one can also trace the second fluxon step with asymptotic velocity $u=2$. It corresponds to the state \state{UDu} and is shown in Fig.~\ref{Fig:IVC:Ud} by \FigIVCuUD. We were able to find this state only starting from the point $e$ at $\gamma=0.3$ with a fluxon situated at $x=5$ and an antifluxon at $x=8$. It is impossible to visualize this mode just by sweeping the bias current since this step is shadowed by \state{u}, \state{UUDd} and \state{Ud} steps.
\begin{figure}[tb]
  \centering\includegraphics{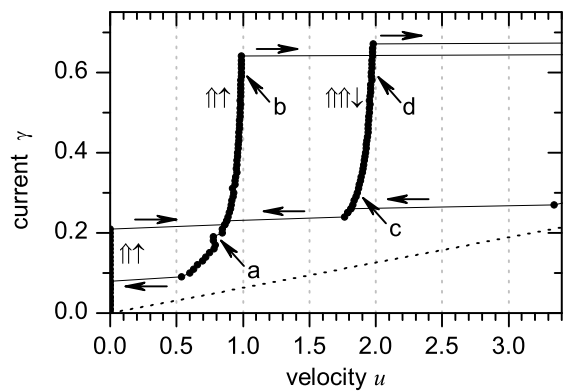}
  \caption{
    Current voltage characteristic after injection of fluxon into 0-$\pi$-LJJ containing a semifluxon of positive polarity. All (even and odd) fluxon steps can be traced. The maximum supercurrent $\gamma_{\max}\approx 0.21$. The trajectories at the bias points a--d are shown in Fig.~\ref{Fig:Traj:Uu}(a)--(d).
  }
  \label{Fig:IVC:Uu}
\end{figure}

\subsubsection{Positive semifluxon}

Now we consider a semifluxon of a positive polarity initially present in the LJJ. The injected fluxon and semifluxon repel each other and no annihilation can occur. 
Here we use the boundary conditions (\ref{Eq:BC:Unconv}) with $k_F=1$ and $k_{\rm SF}=1$ to obtain the IVC. By applying a small bias current we push the fluxon along the ring so it approaches the semifluxon. For a small value of the bias current the situation is static as the driving force of the bias current can be compensated by the repulsion force between the fluxon and the semifluxon. By increasing the bias current, the fluxon moves closer to the semifluxon and at some critical value of the bias current overcomes the maximum possible repulsion force and passes through the semifluxon. After this, the fluxon keeps moving around the ring, bumping the semifluxon once per cycle. The current voltage characteristic of this state is shown in Fig.~\ref{Fig:IVC:Uu}. One can see that the maximum supercurrent $\gamma_c^{\stat{Uu}}=\frac{2}{3\pi}\approx 0.21$ (see Appendix \ref{Sec:AnalDepin}), corresponding to the maximum possible repulsion force between the fluxon and the semifluxon, is considerably smaller than $\gamma_c^\uparrow=2/\pi$ of a single semifluxon. Thus, the two situations can be distinguished.

\begin{figure*}[tb]
  \centering\includegraphics{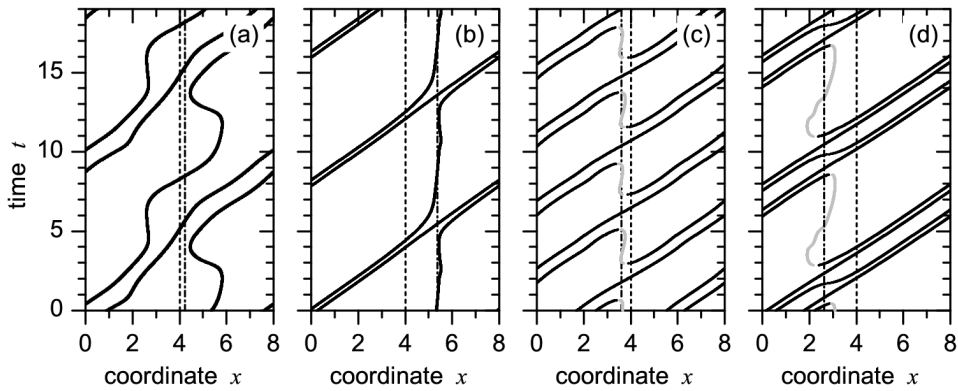}
  \caption{
    Trajectories of semifluxons in the state \state{Uu} (first fluxon step) for $\gamma=0.2$(a) and $\gamma=0.6$ (b) and in the state $\stat{UUd}$ for $\gamma=0.3$ (c) and $\gamma=0.6$ (d). The vertical dashed line shows the position of the discontinuity. The vertical dotted line shows the position of the center of the static semifluxon at the corresponding bias. In (a) for $\gamma=0$ both lines coincide.
  }
  \label{Fig:Traj:Uu}
\end{figure*}

The trajectories corresponding to the dynamics of the system are shown in Fig.~\ref{Fig:Traj:Uu} for several bias points marked in Fig.~\ref{Fig:IVC:Uu}. We see that at the first fluxon step, see Fig.~\ref{Fig:Traj:Uu}(b), the fluxon (double line) moves progressively colliding with the semifluxon. The semifluxon corresponds to a more or less vertical line shifted to the right from the discontinuity point by the bias current. Note, that in Fig.~\ref{Fig:Traj:Uu}(a) we can see essentially the same dynamics, but the semifluxon is much more delocalized, probably because bias point $a$ corresponds to the resonance which can be seen on the IVC in Fig.~\ref{Fig:IVC:Uu}. We believe that this resonance may be related to the eigen-modes of the semifluxon and to the Cherenkov emission of the plasma waves which are excited when the fluxon periodically bumps the semifluxon. The emitted plasma wave forms a standing wave which interacts with a fluxon and results in the resonance on the IVC.

Thus, the polarity of the semifluxon can be probed by inserting a test fluxon of known polarity and measuring the IVC. If a fluxon and a semifluxon have the same polarity one should expect $I_{\max}=\frac{2}{3\pi}\approx 0.21 I_c$ and the appearance of the first fluxon step. If a fluxon and a semifluxon have different polarities, one should expect $I_{\max}= \frac{2}{\pi} I_c\approx 0.63 I_c$ and no first fluxon step provided the injection was made at $\gamma=0$. 

Another difference between these two configurations is the value of the retrapping current $\gamma_r$. As one can see from Figs.~\ref{Fig:IVC:Ud} and \ref{Fig:IVC:Uu} the value of $\gamma_r$ in the $\stat{Ud}=\stat{u}$  state is at least twice larger than for the \state{Uu} state.

An interesting observation can be made, if we trace all fluxon branches of the IVC. In a usual annular LJJ with only one trapped fluxon (without a semifluxon) one can observe fluxon steps at $V=nV_1$ with only odd numbers $n$: the first one appears due to the motion of a single fluxon, the third one corresponds to an additional fluxon-antifluxon pair generated, the fifth one to two fluxon-antifluxon pairs, \etc. 

In an annular LJJ containing a fluxon and a semifluxon of the same polarity the situation is different. One can observe \emph{fluxon steps with any integer $n$}. This is especially easy to observe for the \state{Uu} state. The reason for this becomes clear after analysis of the trajectories shown in Fig.~\ref{Fig:Traj:Uu} for several bias points marked in Fig.~\ref{Fig:IVC:Uu}.

We see that at the second step, there are two positive fluxons moving in the same direction and the semifluxon became of negative polarity \state{UUd}. This means that the semifluxon flipped, and has emitted one positive fluxon --- the process opposite to annihilation. Now, since there are two fluxons which can move and one negative semifluxon which cannot, the asymptotic voltage of the step is equal to $2V_1$. Still, the system can generate fluxon-antifluxon pairs which, together with semifluxon flipping, will result in fluxon steps for all integer $n$.

Actually, a similar effect can be observed when initially the semifluxon had negative polarity. Then starting from the state \state{u} one could trace the fluxon steps corresponding to the states \state{Ud}, \state{UDu}, \state{UUDd} (see Fig.~\ref{Fig:IVC:Ud}). The trajectories for some of the bias points can be seen in Fig.~\ref{Fig:Traj:Ud}.

We also note that when the semifluxon polarity changes, the shift of the average position of the semifluxon changes too. This can be visualized using low temperature scanning electron microscopy\cite{Laub:1995:LTSEM-FluxonContraction}. 

Analyzing the trajectories while the bias point moves along the second fluxon step we discovered that the two fluxons are moving more or less equidistantly at a low value of the bias current, \ie, at the bottom of the step, as can be seen in Fig.~\ref{Fig:Traj:Uu}(c). When the bias current increases up to $\gamma\approx0.45\ldots0.50$ or above, the fluxons bunch together as shown in Fig.~\ref{Fig:Traj:Uu}(d). The bunching appears because the plasma waves, emitted during fluxon-semifluxon collisions, result in an effective attracting potential between two fluxons. We do not discuss this phenomenon in detail here, but note that similar bunching was observed in JJ arrays\cite{Ustinov:JJA:FluxonBunching} or in stacks of LJJs\cite{Goldobin:Cherry1}. In our case, the role of periodic obstacle is played by a semifluxon. Generally speaking, the plasma wave emission has a Cherenkov origin and is simply related to the peculiar dispersion relation for plasma waves in the system under question\cite{Goldobin:Cherry2}.

\subsection{Two semifluxons in a ring}
\label{Sec:2SF}

In the case of more than one semifluxon of unknown polarity, the above read out procedure cannot give information on the polarity of each of the semifluxons, but allows to distinguish between states (a) \state{uu}, (b) \state{dd}, and (c) \state{ud} or \state{du}. In other words, we can find out the total flux hold by semifluxons. Injecting the first fluxon in a similar fashion as above, we measure the IVC. If the critical current is low, the fluxon does not annihilate with semifluxons and keeps moving around the ring. This can happen only if in the initial state both semifluxons had the same polarity as the injected fluxon. If, after injection of a fluxon, $I_c$ is still high, this means that the injected fluxon has flipped one of the semifluxons. In this case we inject a second fluxon and so on. If $I_c$ becomes lower after the first injection, the initial state has been \state{uu}; if $I_c$ becomes lower after second injection, the initial state has been \state{ud} or \state{du}; and if $I_c$ becomes lower after the third injection, the initial state has been \state{dd}.

\section{Conclusions}
\label{Sec:Conclusion}

We have shown that by introducing a test fluxon of known polarity into the LJJ with a semifluxon of unknown polarity we can destructively read out the semifluxon state. In principle, the read out can be made non-destructive, if after reading out the semifluxon state we introduce another fluxon of opposite polarity so that the system returns to its initial state (as before read out) as a result of fluxon-fluxon or fluxon-semifluxon annihilation. In case of two or more semifluxons one can read out the total number of positive and negative semifluxons. This technique is also applicable to arbitrary fractional $\kappa$-vortices, but not very close to $\kappa=0$ or $\kappa=2\pi$ where depinning currents for both $N=0$ and $N=1$ approach zero.

We have also investigated fluxon steps in an annular LJJ with a $\pi$-discontinuity. We found that one can trace the fluxon steps corresponding to \emph{all} integer $n$, instead of only odd $n$ like in conventional LJJ with one trapped fluxon. We have observed a smooth transition to the state of two bunched fluxons in the \state{UUd} state.

Further, in appendix \ref{Sec:AnalDepin} we have derived the Eq.~(\ref{Eq:gamma_c}) which gives the depinning current of $N$ fluxons pinned by an arbitrary $\kappa$-vortex. We discovered that the biggest obstacle for a fluxon is a fractional $\kappa_1$-vortex with $\kappa_1\approx0.861\pi$ rather than a semifluxon. The formula (\ref{Eq:kappa_n}) allows to compute the size $\kappa_N$ of the fractional vortex which is the biggest obstacle for $N$ fluxons trying to pass it.

\begin{acknowledgments}
  E.G. thanks H. Susanto and S. van Gils for fruitful discussions and hospitality during his visit to University of Twente.
  This work was supported by the Deutsche Forschungsgemeinschaft, and by the ESF programs "Vortex" and "Pi-shift".
\end{acknowledgments}

\appendix

\section{Depinning of fluxons by an arbitrary fractional vortex}
\label{Sec:AnalDepin}

Since one can study experimentally arbitrary $\kappa$-vortices\cite{Goldobin:Art-0-pi}, here we calculate the depinning current for a chain of $N$ fluxons that are pinned by and are trying to pass an arbitrary $\kappa$-vortex in a LJJ of infinite length.

The static version of Eq.~(\ref{Eq:sG-mu}) for arbitrary discontinuity $\kappa$ is:
\begin{equation}
  \mu_{xx} = \sin[\mu-\theta(x)]-\gamma
  , \label{Eq:mu:static}
\end{equation}
where 
\[
  \theta(x)=\left\{
  \begin{array}{ll}
    0 & \mbox{for }x<0,\\
    -\kappa & \mbox{for }x>0
  \end{array}
  \right.
\]

We write separately the Eqs. for the part of the junction to the left and to the right from discontinuity, situated at $x=0$. After integration one arrives to the following two equations.

\begin{subequations}
  \begin{eqnarray}
    y_1(\mu) = \mu_x &=& \pm2\sqrt{C_1(\gamma)-\cos\mu-\gamma\mu}
    ,\quad x<0; \label{Eq:mux_1}\\
    y_2(\mu) = \mu_x &=& \pm2\sqrt{C_2(\gamma)-\cos(\mu-\kappa)-\gamma\mu}
    ,\quad x>0, \label{Eq:mux_2}
  \end{eqnarray}
  \label{Eq:mux}
\end{subequations}

Assuming that
\begin{eqnarray}
  \mu_x(\pm\infty) &=& 0
  ; \label{Eq:mu_x_1}\\
  \mu(-\infty) &=& \arcsin\gamma
  ; \label{Eq:mu+}\\
  \mu(+\infty) &=& \arcsin\gamma + 2\pi N + \kappa
  , \label{Eq:mu-}
\end{eqnarray}

we arrive to the following expressions for $C_1$ and $C_2$:
\begin{eqnarray}
  C_1(\gamma)&=&\sqrt{1-\gamma^2}+\gamma\arcsin\gamma
  ; \label{Eq:C_1}\\
  C_2(\gamma)&=&\sqrt{1-\gamma^2}+\gamma\arcsin\gamma+(2\pi N+\kappa)\gamma
  , \label{Eq:C_2}
\end{eqnarray}
\begin{figure}[!htb]
  \centering\includegraphics{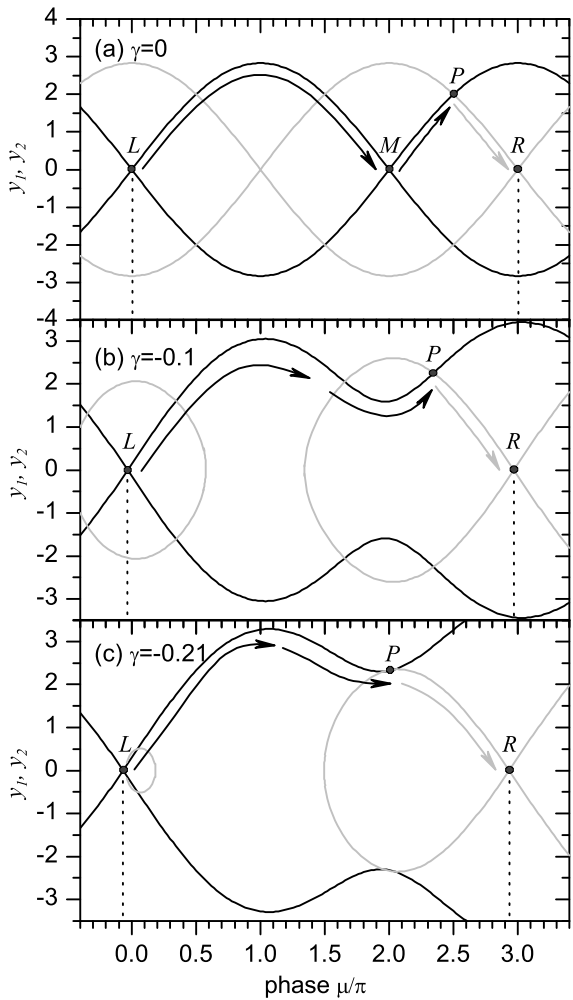}
  \caption{%
    The trajectories on the phase plane corresponding to Eqs.~(\ref{Eq:mux}). Black color corresponds to $y_1$ (0-part), the gray color corresponds to $y_2$ ($\pi$-part). The trajectories are presented for a fluxon pinned by a semifluxon ($N=1$, $\kappa=\pi$) and (a) $\gamma=0$, (b) $\gamma=-0.1$ and (c) $\gamma=0.21$. Arrows are shown to guide the eye. Arrows show the path which corresponds to going from $x=-\infty$ (point $L$) to $x=+\infty$ (point $R$).
  }
  \label{Fig:PhasePlane}
\end{figure}

We use the phase plane analysis\cite{Susanto:SF-gamma_c} to find possible static configurations. The trajectories on the phase plane $y_{1,2}(\mu)$ corresponding to Eqs.~(\ref{Eq:mux}) are shown in Fig.~\ref{Fig:PhasePlane} for $\gamma=0$, $\gamma<\gamma_c$ and $\gamma=\gamma_c$. At $\gamma=0$ the fluxon and fractional vortex are separated by a large distance. The fluxon corresponds to the trajectory between point $L$ ($x=-\infty$) and point $M$. The fractional vortex corresponds to the trajectory between point $M$ and point $R$ ($x=+\infty$) and contains a point $P$ where the black and gray trajectories intersect, \ie, the phase crosses the 0-$\pi$-boundary. From Fig.~\ref{Fig:PhasePlane}(a)--(c) one can see that with increasing $\gamma$ the intersection point $P$ shifts until, at the critical value of $\gamma=\gamma_c$ the trajectories just touch each other at point $P$, as shown in Fig.~\ref{Fig:PhasePlane}(c). For larger $\gamma$, no intersection is possible and the static solution does not exist.

The critical value of $\gamma$ at which the switching between trajectories is still possible is defined by the following conditions:
\begin{eqnarray}
  y_1(\mu)=y_2(\mu)
  ; 
  y'_1(\mu)=y'_2(\mu)
  , 
\end{eqnarray}
This conditions are satisfied for $\mu=\frac{1}{2}(3\pi+\kappa)$, which leads us to the final result:
\begin{equation}
  \gamma_c(\kappa)=\frac{2}{2\pi N+\kappa}\sin\frac{\kappa}{2}
  . \label{Eq:gamma_c}
\end{equation}
\begin{figure}[!htb]
  \centering\includegraphics{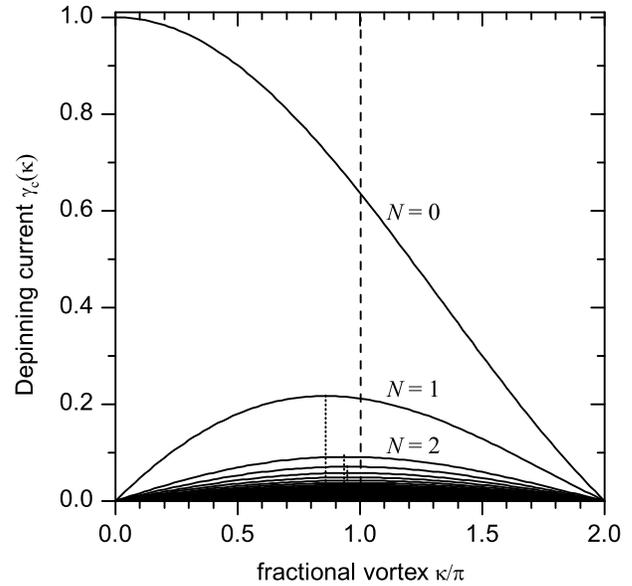}
  \caption{%
    The dependence $\gamma_c(\kappa)$ calculated for different $N$ using Eq.~(\ref{Eq:gamma_c}). The dashed line shows $\kappa=\pi$ corresponding to the semifluxon. Three dotted lines show the values $\kappa_1$, $\kappa_2$ and $\kappa_3$ which correspond to the fractional vortices which cause the maximum possible pinning for $N=1,2,3$ fluxons.
  }
  \label{Fig:Ic(kappa)}
\end{figure}

The plot of this dependence for different $N$ is shown in Fig.~\ref{Fig:Ic(kappa)}. The $\gamma_c(\kappa)$ has reasonable limiting behavior (absence of pinning for $N>0$) for $\kappa\to0$ and $\kappa\to 2\pi$. For depinning of a fluxon ($N=1$) by a semifluxon, $\gamma_c(\pi)=2/3\pi\approx 0.21$. It is interesting that $\gamma_c(\kappa)$
has a maximum not at $\kappa=\pi$ but a bit shifted, at $\kappa_1=0.861\pi$, \ie, a $\kappa_1$-vortex is the biggest obstacle for a fluxon. For two and three fluxons the biggest obstacles are vortices with $\kappa_2=0.918\pi$ and $\kappa_3=0.942\pi$. All $\kappa_N$ can be found from the equation
\begin{equation}
  \tan\frac{\kappa_N}{2} = \frac{\kappa_N}{2}+N\pi
  . \label{Eq:kappa_n}
\end{equation}
Note, that in the limit $N\to\infty$, $\kappa_N\to\pi$.

Our result (\ref{Eq:gamma_c}) is also valid in the case $N=0$ and gives the ``depinning'' current of the $\kappa$-vortex itself. The result (\ref{Eq:gamma_c}) coincides with the expression obtained recently for the critical current of an annular junction in the presence of a $\kappa$-discontinuity of the phase (current dipole) created by injectors\cite{Ustinov:2002:ALJJ:Ic(Iinj)}. In fact one can even say that our result (\ref{Eq:gamma_c}) and the one of Ref.~\onlinecite{Ustinov:2002:ALJJ:Ic(Iinj)} are \emph{the same} at least if one considers annular LJJ. In fact, if we denote $\kappa'=2\pi N+\kappa$, we can rewrite (\ref{Eq:gamma_c}) as
\begin{equation}
  \gamma_c(\kappa')=\frac{2\sin\frac{\kappa'}{2}}{\kappa'}
  , \label{Eq:Ic-ann}
\end{equation}
exactly as in Ref.~\onlinecite{Ustinov:2002:ALJJ:Ic(Iinj)}.
Physically this means that when one creates a large discontinuity $\kappa'=2\pi N+\kappa$ in an annular LJJ, it automatically relaxes into $N$ fluxons trying to pass through the fractional $\kappa$-vortex under the action of bias current. The interesting result is that the Eq.~(\ref{Eq:gamma_c}) for an infinitely long linear junction and Eq.~(\ref{Eq:Ic-ann}) for an annular one coincide regardless of ``fluxon crowding'' which may take place in annular LJJ because of its finite length.

\bibliography{LJJ,pi,SF,QuComp,SFS,jj-arrays}

\end{document}